\documentclass[printnumbers,amsmath,amssymb,twocolumn]{revtex4-2}
\usepackage{graphicx}
\usepackage{color}
\usepackage{float}
\usepackage{epstopdf}
\usepackage{hyperref}

\begin{document}

\title{Universal link between structural relaxation and self-diffusion  in simple supercooled liquids}

 \author{ Mingyu Zhu$^{1}$, Xinyu Chang$^{1}$, Licun Fu$^{1}$, Yiming Zheng$^{1}$, and Lijin Wang$^{1,*}$ \\
$^{1}${School of Physics, State Key Laboratory of Opto-Electronic Information Acquisition and Protection Technology,  Anhui University, Hefei 230601, China}\\
$^{*}$ Corresponding author,  Email: lijin.wang@ahu.edu.cn\\
}

\begin{abstract}

Despite  a vigorous effort,  a comprehensive and satisfactory understanding of  the dynamics underlying the glass transition  in finite-dimensional systems has not  yet  emerged. One key  hallmark  of   supercooled liquids approaching the glass transition is the breakdown of the classical Stokes–Einstein (SE) relation, manifested as the decoupling between the self-diffusion coefficient and  structural relaxation time, commonly described by a fractional SE relation.
Whether and how the fractional SE  exponents vary remains an ongoing debate, even for  the simplest supercooled liquids.  In this work, we numerically establish a universal relation between the self-diffusion coefficient and  structural relaxation time by introducing characteristic  scales that are revealed under the condition of constant dynamical heterogeneity. Moreover, this universal relation could also capture literature data for  supercooled  water and silica. Remarkably, we find that the fractional SE exponent is invariant across various  supercooled liquids studied spanning a wide range of fragility, thereby challenging the prevailing view that the fractional SE exponent is  correlated with fragility. Our study demonstrates that investigating dynamical heterogeneity yields more valuable insights into understanding glassy dynamics than  expected. 
\end{abstract}

\date{\today}

\maketitle

\section*{Introduction}

\noindent  When subjected to sufficiently high cooling rates, a liquid can bypass crystallization, first forming a supercooled liquid below its  melting temperature and ultimately transforming into an amorphous glassy state upon crossing the glass transition temperature. Despite extensive research efforts, a comprehensive and satisfactory theoretical description of the glass transition remains elusive~\cite{Dyre_2026_NRP,Royall_2024_RMP,Berthier_2011_RMP,Debenedetti_2001_Nature,Tanaka_2010_NM}.
On approaching the glass transition, the dynamics of supercooled liquids  slows down dramatically with slight decrease in temperature. In general, the  slowdown is more pronounced in fragile supercooled liquids than in strong ones~\cite{Berthier_2011_RMP,Angell_1991_JNCS,Angell_1995_Science}. For strong supercooled liquids, the temperature dependence of the structural relaxation time $\tau$ obeys  the Arrhenius law.  In contrast, fragile supercooled liquids typically deviate from the Arrhenius law—the more fragile the liquid, the greater the deviation.

For  high-temperature liquids,  the self-diffusion coefficient $D$ and structural relaxation time  are related through   the classical Stokes-Einstein  (SE) relation,  $D \sim (\tau/T)^{-1}$~\cite{Oppelstrup_2025_JCP,Einstein_SE}.  Nevertheless,  the SE relation breaks down at low temperatures in supercooled liquids, regardless of whether they are strong or fragile~\cite{Berthier_2011_RMP,Fujara_1992_ZPB_EdigerPRL,Li_2019_PNAS, Ediger_2000_AnnuRev,Sengupta_2013_JCP,Adhikari_2021_JPCB,Debenedetti_2001_Nature,Staley_2015_JCP,Ediger_2000_AnnuRev,Glotzer_2000_JCP,Chen_2006_PNAS,Flenner_2013_JCP,Xu_2019_NP,Dehaoui_2015_PNAS,Marcorini_2025_JCP,Henritzi_2015_JCP,Swallen_2003_PRL,Becker_2006_PRL,Flenner_2014_PRL, Pan_2017_SR}. 
Specifically,  it is well established  that the self-diffusion coefficient exhibits an enhancement by several orders of magnitude relative to the value expected from the structural relaxation time via the SE relation, thereby indicating a pronounced decoupling between  the self-diffusion coefficient and structural relaxation time.    Additionally,  experimental studies on supercooled colloidal fluids have also demonstrated a breakdown of the SE relation~\cite{Edmond_2012_PNAS,Mishra_2015_PRL}.  
Diverse explanations have been put forward to account for the breakdown, yet no consensus has  emerged~\cite{Kumar_2006_JCP,Starr_2013_JCP,Ediger_2000_AnnuRev,Berthier_2004_PRE,Berthier_2011_RMP,Tarjus_1995_JCP,Charbonneau_2014_PNAS,Hodgdon_1993_PRE,Glotzer_2000_JCP,Xu_2019_NP,Jung_2004_PREjcp,Pan_2005_PREjcp,Xia_2001_JPCBjcp,Biroli_2007_JPCMjcp,Chong_2008_PREjcp,Diezemann_1998_PREjcp,Ngai_1998_JPCBjcp,Zangi_2007_PREjcp,Henritzi_2015_JCP,Parmar_2017_PRL,Sengupta_2013_JCP,Mishra_2015_PRL}.  A commonly accepted  perspective~\cite{Starr_2013_JCP,Ediger_2000_AnnuRev,Berthier_2004_PRE,Berthier_2011_RMP,Glotzer_2000_JCP,Tarjus_1995_JCP,Jung_2004_PREjcp,Pan_2005_PREjcp,Henritzi_2015_JCP,Parmar_2017_PRL,Starr_2013_JCP,Sengupta_2013_JCP,Flenner_2014_PRL,Xia_2001_JPCBjcp} is that the breakdown is a consequence of the emergence of dynamical heterogeneity~\cite{Jung_2023_PRL,Berthier_2011_RMP,Ediger_2000_AnnuRev,Richert_2002_JPCM,Chacko_2021_PRL,Ozawa_2023_PRL,Tahaei_2023_PRX,Berthier_2011_Oxford}, wherein   some regions temporarily exhibit  higher-than-average particle mobility and others slower motion. In broad terms, this perspective holds that  the  diffusion and  relaxation are predominantly governed by fast and slow particles, respectively.

Empirically, the decoupling  between the self-diffusion coefficient and structural relaxation time could be  described by  a  fractional SE relation,  $D \sim (\tau/T)^{-\xi}$,  where the fractional SE exponent $\xi$ is smaller than $1$ and measures the extent of the SE relation breakdown.    Whether and how $\xi$  varies  has attracted a great interest.  Theoretically, the dynamical facilitation model and   random first-order transition theory both made a prediction that $\xi$  correlates  with fragility~\cite{Pan_2005_PREjcp,Xia_2001_JPCBjcp,Jung_2004_PREjcp}. Note that fragility is greater in more fragile liquids. This prediction is corroborated by several numerical simulations~\cite{Adhikari_2021_JPCB,Ozawa_2016_JSM} and experimental observations~\cite{Ediger_2000_AnnuRev}.
However, a simulation study~\cite{Bell_2020_NC} reported that $\xi \approx 0.73$ for   strongly correlating liquids characterized by pronounced virial and potential-energy correlation~\cite{Dyre_2014_JPCB}, suggesting that, in such systems, $\xi$ may be invariant with respect to variations in fragility.
Nevertheless, it remains unclear whether this result can be generalized to weakly correlating  liquids.  More recently, numerical investigations~\cite{Qin_2023_PRB}  indicated that $\xi$ decreases with increasing fragility, which is consistent with theoretical predictions~\cite{Pan_2005_PREjcp,Xia_2001_JPCBjcp}; however, it eventually reaches a saturation value once the fragility surpasses a certain threshold.  Additionally,  other pertinent investigations did not yield convergent conclusions~\cite{Chen_2006_PNAS,Becker_2006_PRL}. 
Therefore, a  universal consensus  still seems  lacking regarding how the fractional SE exponent varies. 

One compelling and crucial research direction is to  determine whether a universal relationship exists between the self-diffusion coefficient and  structural relaxation time across a broad class of supercooled liquids. Establishing such a universal correlation would provide an alternative framework for resolving the puzzle associated with the fractional SE exponent. In particular, if datasets of the self-diffusion coefficient plotted against the structural relaxation time for different systems can be collapsed onto a single master curve through appropriate rescaling, then the fractional SE exponent can be obtained directly from the scaling behavior of this master curve and is, by construction, unique.

Flenner \textit{et al}.~\cite{Flenner_2014_PRL} demonstrated the possibility of obtaining a universal correlation  by rescaling the self-diffusion coefficient and  structural relaxation time with  appropriate scaling factors in several binary model systems. However, the physical meaning of   scaling factors adopted therein remains poorly understood. One subsequent  study by Bell \textit{et al}.~\cite{Bell_2020_NC}  attempted to rationalize the universality observed by  Flenner \textit{et al}.~\cite{Flenner_2014_PRL} by applying  an excess-entropy scaling strategy originally  introduced  by Rosenfeld~\cite{excess_Rosenfeld_1977, excess_Dyre_2018}.  Bell \textit{et al}.~\cite{Bell_2020_NC} identified a universal correlation between the self-diffusion coefficient and viscosity $\eta$ in strongly correlating binary  supercooled liquids. Note that the viscosity is approximately  proportional to the structural relaxation time,  as dictated by the Maxwell’s relation. However, the excess-entropy scaling framework fails to accurately describe the behavior of weakly correlated supercooled liquids, particularly in systems where the constituent particles interact via soft-core potentials~\cite{Pond_2011_SM,Krekelberg_2009_PRE,Gnan_2009_JCP}.
Therefore,  it remains  unclear whether a universal correlation exists between the self-diffusion coefficient and structural relaxation time across a wider range of  supercooled liquids (encompassing both strongly and weakly correlating systems), and if such a  correlation exists, how to establish it reasonably.   

In this work, we perform numerical investigations on a broad class of model supercooled liquids with markedly different  interaction potentials. We demonstrate that the dynamical heterogeneity provides a key framework for establishing a universal relationship between the self-diffusion coefficient and  structural relaxation time. In particular, we show that imposing the condition of constant dynamical heterogeneity is crucial for uncovering this universality.
Under this constraint, we identify two system-dependent characteristic scales, $D^*$ and $(\tau/T)^*$. Upon rescaling $D$ and  $\tau/T$ by $D^*$ and $(\tau/T)^*$, respectively, we obtain an excellent scaling collapse for all supercooled liquids examined,  including both strongly correlating and weakly correlating  liquids. Furthermore, the master curve generated by this data collapse can incorporate additional literature data for complex supercooled liquids.
This universal representation enables a more robust determination of the fractional SE exponent by fitting the master curve to the functional form, $D \sim (\tau/T)^{-\xi}$. Two distinct scaling regimes emerge: at high temperatures, $\xi = 1$, consistent with the  SE relation; at low temperatures, a fractional SE relation is observed with $\xi \approx 0.728$. This result indicates a universal breakdown of the SE relation, wherein the fractional SE exponent  remains invariant with respect to  system-specific properties, such as fragility, across all supercooled liquids considered. Our study highlights the central role of dynamical heterogeneity in understanding glassy dynamics.

\section*{Results}

\subsection*{Preparing  systems}

We prepare systems with significantly distinct properties, such as fragility, by adjusting  interaction potentials or/and pressures.  The interaction potentials  considered include the harmonic (Harm) potential,  Hertzian (Hertz) potential~\cite{Wang_2012_SM}, inverse-power-law (IPL) potential~\cite{Ninarello_2017_PRX}, and Lernnard-Jonnes (LJ) potential~\cite{Kob_1994_PRL}. Additionally, we  study  $\rm{Zr}_{50}\rm{Cu}_{50}$ metallic supercooled  liquids where atoms interact via the realistic embedded atom method (EAM) potential~\cite{Cheng_2009_EAM}.  Systems  with  the IPL potential are polydisperse mixtures, whereas all other investigated systems with Harm, Hertz,  LJ, and  EAM potentials are binary mixtures.

We perform three-dimensional molecular dynamics simulations, using LAMMPS~\cite{lammps}, in the \textit{NPT} ensemble to generate dynamical trajectories. The structural relaxation time $\tau$ is determined as the time  when the self-intermediate scattering function decays to a value of $\rm{1/e}$.  The self-diffusion coefficient $D$ is obtained by fitting the long-time mean squared displacement using the Einstein relation.  Dynamical heterogeneity is commonly  characterized  by the non-Gaussian parameter $\alpha_2(t)$ or the four-point dynamical susceptibility $\chi_4(t)$~\cite{Glotzer_2000_JCP,Kim_2013_JCP,Kob_1997_PRL_A2,Henritzi_2015_JCP,Starr_2013_JCP,Adhikari_2021_JPCB,Sengupta_2013_JCP}. In the main text, we primarily present  results obtained from the non-Gaussian parameter.   $\alpha_2(t)$ typically exhibits a non-monotonic dependence on time $t$  in supercooled liquids. We adopt  the maximum of $\alpha_2(t)$,  $\alpha_{2,\mathrm{max}}$, to quantify the strength of dynamical heterogeneity.  We have also verified that our major conclusions remain unchanged when the four-point dynamical susceptibility is employed to characterize dynamical heterogeneity (see the Supplementary Information). 
More simulation details are provided in the Methods section.

\begin{figure}[t]
\includegraphics[width=0.48\textwidth]{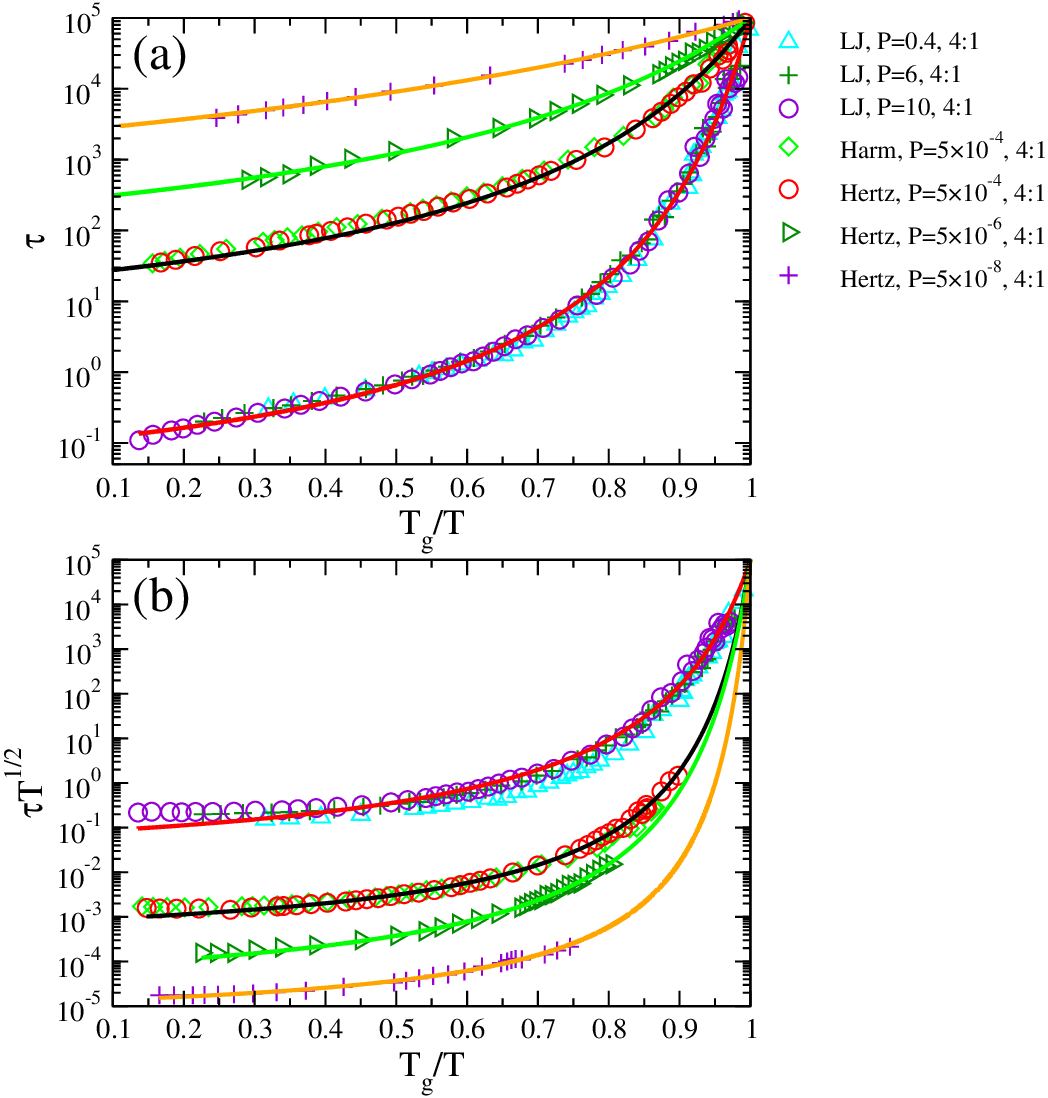}
\caption{\label{fig:fragility} (a) Typical Angell plots of structural relaxation time $\tau$ versus scaled reciprocal temperature $T_{\rm g}/T$ for binary (4:1) systems with Harm,  Hertz, and LJ potentials at different pressures. The glass transition temperature $T_{\rm g}$  is defined as  $\tau(T_{\rm g})=10^{5}$. (b) The other type of Angell plots,  $\tau T^{1/2}$ versus  $T_{\rm g}/T$. Here, $T_{\rm g}$ is defined as the temperature at which  $\tau T^{1/2}=10^{5}$. The solid curves represent Vogel-Fulcher-Tammann
 fits~\cite{Berthier_2011_RMP}, by means of which we can precisely locate  $T_{\rm g}$.   Note that all systems shown here have the same units of length, mass, and energy (see Methods section for more details). The ratio 4:1 in the legend corresponds to a 4:1 composition ratio in  binary mixtures.  Symbols in (a) and (b) represent the same meanings.
}
\end{figure}

One of our major objects in this study is to check whether the fractional SE exponent changes with fragility, and if so, how it changes.  Therefore, we first demonstrate that our systems under study exhibit different fragility. To compare  fragility, we present in Fig.~\ref{fig:fragility}  two types of Angell plots~\cite{Berthier_2011_RMP,Angell_1991_JNCS,Angell_1995_Science}  for systems with Harm,  Hertz, and LJ potentials at different pressures.   
Shown in Fig.~\ref{fig:fragility}(a) is the  typical Angell plots, i.e., the plot of $\tau$ versus $T_{\rm{g}}/T$, where $T_{\rm {g}}$ is the glass transition temperature whose definition is in the caption of Fig.~\ref{fig:fragility}. 
Following Ref.~\cite{Massimo_2020_PRL}, we present in Fig.~\ref{fig:fragility}(b) the other  type of Angell plots  to ensure dimensional consistency, specifically, the plot of $\tau T^{1/2}$ versus $T_{\rm{g}}/T$.  Typically,  a straight line in Angell plots indicates the Arrhenius behavior and thus represents strong supercooled liquids, whereas a curve indicates the deviation from the Arrhenius behavior and hence  represents fragile supercooled liquids, i.e., a steeper curve corresponds to more fragile supercooled liquids~\cite{Berthier_2011_RMP,Angell_1995_Science}.   In both types of Angell plots in Fig.~\ref{fig:fragility},  systems with distinct Angell plots and hence different fragility could be observed. Notice that we do not aim to assert which system exhibits greater fragility, as  the two types of plots seem to yield opposite conclusions on this matter. Instead, we assert that the systems under investigation exhibit distinct fragility, independently of the two aforementioned approaches used to characterize fragility.

\subsection*{Characteristic scales under constant dynamical heterogeneity condition}

The fractional SE relation in supercooled liquids describes how  $D$  is linked to $\tau$.   We will demonstrate   that the reduced  $D$  is a univariate function of  dynamical heterogeneity, and the same holds true for the reduced $\tau/T$. Therefore, a universal connection between $D$  and $\tau/T$ can be established with the dynamical heterogeneity acting as a bridge.

\begin{figure}[t]
\includegraphics[width=0.48\textwidth]{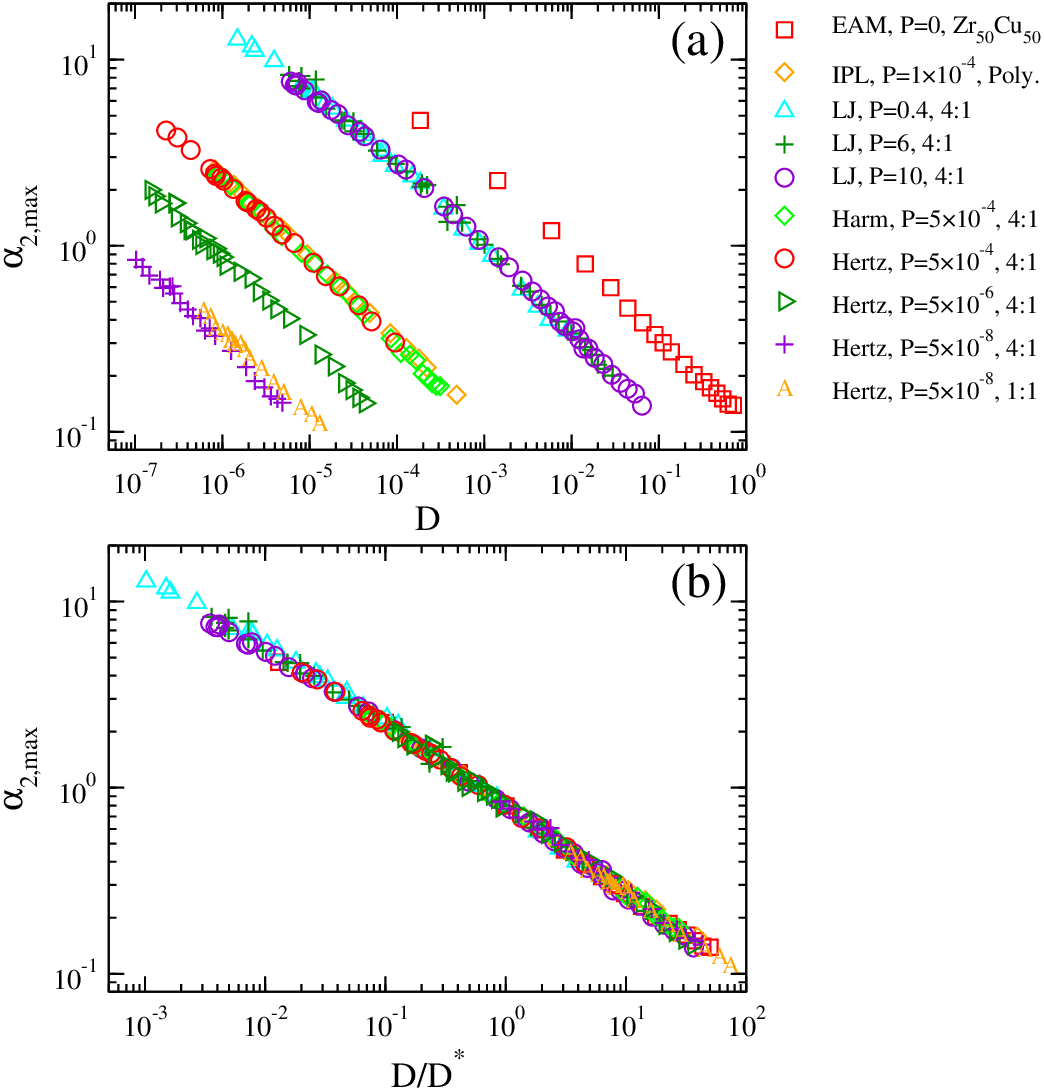}
\caption{\label{fig:D} (a) Correlation between dynamical heterogeneity characterized by the maximum non-Gaussian parameter $\alpha_{\rm 2,max}$, and the self-diffusion coefficient $D$ for  all investigated systems, including 1:1  and 4:1 binary mixtures as well as polydisperse mixtures. (b) Scaling collapse between $\alpha_{\rm 2,max}$ and $D$.  Here,  the characteristic sale $D^{*}$ is  system-dependent, and determined as when  $\alpha_{\rm 2,max}(D)\approx 0.8$. }
\end{figure}

First, we describe the procedure by which we establish a relationship between the diffusion coefficient  and dynamical heterogeneity quantified by  \(\alpha_{\mathrm{2,max}}\) across a broad range of systems. Shown in 
Fig.~\ref{fig:D}(a) is how $\alpha_{\rm{2,max}}$ varies with   $D$   in all systems investigated. Apparently, the  $\alpha_{\rm{2,max}}$   versus $D$ curves differ a lot among most systems. What's  common is that  $\alpha_{\rm{2,max}}$  increases as  $D$ decreases in all studied systems, thus suggesting the enhancement of dynamical heterogeneity upon approaching the glass transition~\cite{Kob_1997_PRL_A2,Berthier_2011_RMP,Ediger_2000_AnnuRev}. 

For a long time,  the establishment of a universal scaling between the self-diffusion coefficient and other properties has attracted considerable interest~\cite{Ning_2019_PRL,Dzugutov_1996_Nature,Zhang_Liu_2026_NC,excess_Dyre_2018, excess_Rosenfeld_1977, Bell_2020_NC,Pond_2011_SM,Krekelberg_2009_PRE,Gnan_2009_JCP}. 
One example is that, within the framework of excess-entropy scaling~\cite{Bell_2020_NC,Pond_2011_SM,Gnan_2009_JCP}, the dimensionless self-diffusion coefficient in strongly correlating supercooled liquids (sharing an identical density-scaling exponent~\cite{Dyre_2014_JPCB}) has been reported to constitute a unique function of the excess entropy.
Nevertheless, this strategy cannot be generalized to describe weakly correlating systems~\cite{Pond_2011_SM,Krekelberg_2009_PRE,Gnan_2009_JCP}, such as systems with soft-core (Harm and Hertz) potentials we are studying.

 \begin{figure}[t]
\includegraphics[width=0.48\textwidth]{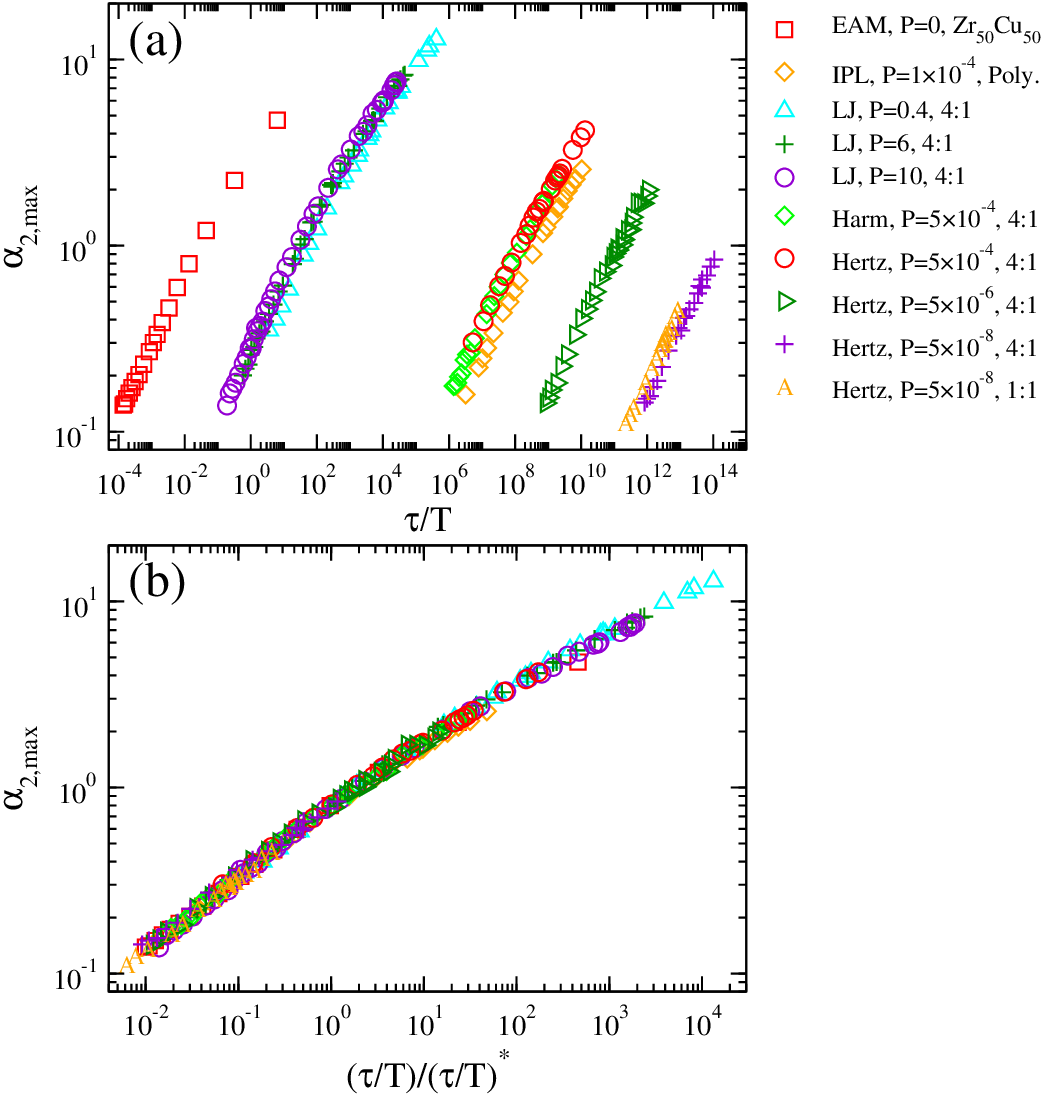}
\caption{\label{fig:tauOverT} (a) Correlation between dynamical heterogeneity characterized by the maximum non-Gaussian parameter $\alpha_{\rm 2,max}$, and the ratio of structural relaxation time to temperature $\tau/T$. (b) Scaling collapse between $\alpha_{\rm 2,max}$ and   $\tau/T$.  The system-dependent characteristic scale $(\tau/T)^{*}$ corresponds to when $\alpha_{\rm 2,max}(\tau/T)\approx 0.8$.
}
\end{figure}

Building upon our previous work~\cite{Wang_2018_PRL,Ren_2020_Scripta}, where we demonstrated that specific characteristic time scales are obscured under the condition of constant dynamical heterogeneity, we extend this framework to examine whether additional characteristic scales required  can likewise be identified under the same condition. Our  analysis shows that, within this constraint, a system-dependent characteristic  scale, denoted by $D^{*}$, can indeed be defined. Utilizing $D^{*}$, we could establish a universal scaling relation linking the self-diffusion coefficient to dynamical heterogeneity,  as demonstrated in Fig.~\ref{fig:D}(b). When $\alpha_{\rm 2,max}$ is plotted against $D/D^{*}$, the data for all investigated systems collapse nicely onto a single master curve. Therefore,  
\begin{equation}
\label{eq:D-alpha2max}
\alpha_{2,\text{max}}= F\left(D/D^*\right).
\end{equation}
 Here, $D^{*}$ in each system is determined such that  $\alpha_{\rm{2,max}}(D^{*})\approx 0.8$. Specifically, for systems where the minimum attainable value of $\alpha_{\rm 2,max}$ is less than 0.8, and the corresponding maximum attainable value exceeds 0.8, we define $D^{*}$ as the value of $D$ at which $\alpha_{\rm 2,max}(D) \approx 0.8$. Plotting   $\alpha_{\rm{2,max}}$ against  $D/D^{*}$ for such systems could yield a single collapsed  curve. However, for systems where the minimum attainable value of  $\alpha_{\rm{2,max}}$  exceeds 0.8,   we horizontally shift the raw  $\alpha_{\rm{2,max}}$ versus $D$  curves (while retaining all $\alpha_{\rm{2,max}}$ values) respectively to align with the aforementioned collapsed curve; the corresponding scaling factor  employed  for this shift in each system is defined as $D^{*}$.    Therefore, under the  condition of constant $\alpha_{\rm{2,max}}$, we demonstrate that the dimensionless self-diffusion coefficient is a single-valued function of $\alpha_{\rm{2,max}}$.  

The original form for the SE relation is $D \sim (\eta/T)^{-\xi}$~\cite{Oppelstrup_2025_JCP,Einstein_SE}. However, it is frequently computationally or experimentally challenging to calculate the viscosity precisely~\cite{Sengupta_2013_JCP}. Therefore,  two variants of the SE relation are commonly utilized in the literature to avoid the calculation of viscosity~\cite{Li_2019_PNAS, Ediger_2000_AnnuRev,Sengupta_2013_JCP,Adhikari_2021_JPCB,Debenedetti_2001_Nature,Staley_2015_JCP,Ediger_2000_AnnuRev,Glotzer_2000_JCP,Chen_2006_PNAS,Flenner_2013_JCP,Xu_2019_NP,Dehaoui_2015_PNAS,Marcorini_2025_JCP,Henritzi_2015_JCP,Swallen_2003_PRL,Becker_2006_PRL,Flenner_2014_PRL, Pan_2017_SR}. The first one is $D \sim (\tau/T)^{-\xi}$; the second one is $D \sim \tau^{-\xi}$. The rationale  behind   the two variants could be found  in Ref.~\cite{Sengupta_2013_JCP}. We aim to incorporate as much as possible  literature data regarding the SE relation. And, to our knowledge,  existing literature contains a larger number of studies that adopt the first variant compared to those that employ the second one. Therefore, we conduct a systematic investigation on the more studied form $D \sim (\tau/T)^{-\xi}$.  Note that we have checked that our major conclusions could  still hold when using the other form.   

We have learned how $D$ correlates with $\alpha_{\rm{2,max}}$ from Fig.~\ref{fig:D}. 
Now we proceed to investigate the link between $ \tau/T$  and $\alpha_{\rm{2,max}}$. Pronounced variations in the dependence of $\alpha_{\rm{2,max}}$ on $\tau/T$ could be observed across  different systems, as presented in Fig.~\ref{fig:tauOverT}(a).
Like what we have done to determine $D^{*}$ in Fig.~\ref{fig:D} under the condition of $\alpha_{\rm{2,max}}\approx 0.8$, we adopt   the same procedure to determine another system-specific characteristic scale  $ (\tau/T)^{*}$.  Notably,  Figure~\ref{fig:tauOverT}(b) demonstrates that results for all systems investigated could collapse onto a single master curve when plotted as $\alpha_{\rm{2,max}}$ against $(\tau/T)/(\tau/T)^{*}$, i.e.,
\begin{equation}
\label{eq:tauOverT-alpha2max}
\alpha_{2,\text{max}}= G\left[(\tau/T)/(\tau/T)^*\right].
\end{equation}
Therefore, we obtain a universal correlation between  $ \tau/T$  and $\alpha_{\rm{2,max}}$ among all systems investigated.  

Equations~\ref{eq:D-alpha2max} and ~\ref{eq:tauOverT-alpha2max} thus suggest that the dynamical heterogeneity  serves as a bridge connecting the two critical parameters in the SE relation: the structural relaxation time and  the self-diffusion coefficient.  Note that this finding holds for all  simple supercooled liquids we studied, including  binary and polydisperse mixtures, along with systems featuring different interaction potentials. 
 Hereafter,  we  demonstrate that  the   SE relation breaks down in a unified manner in all these systems. 

 \begin{figure}[t]
\includegraphics[width=0.48\textwidth]{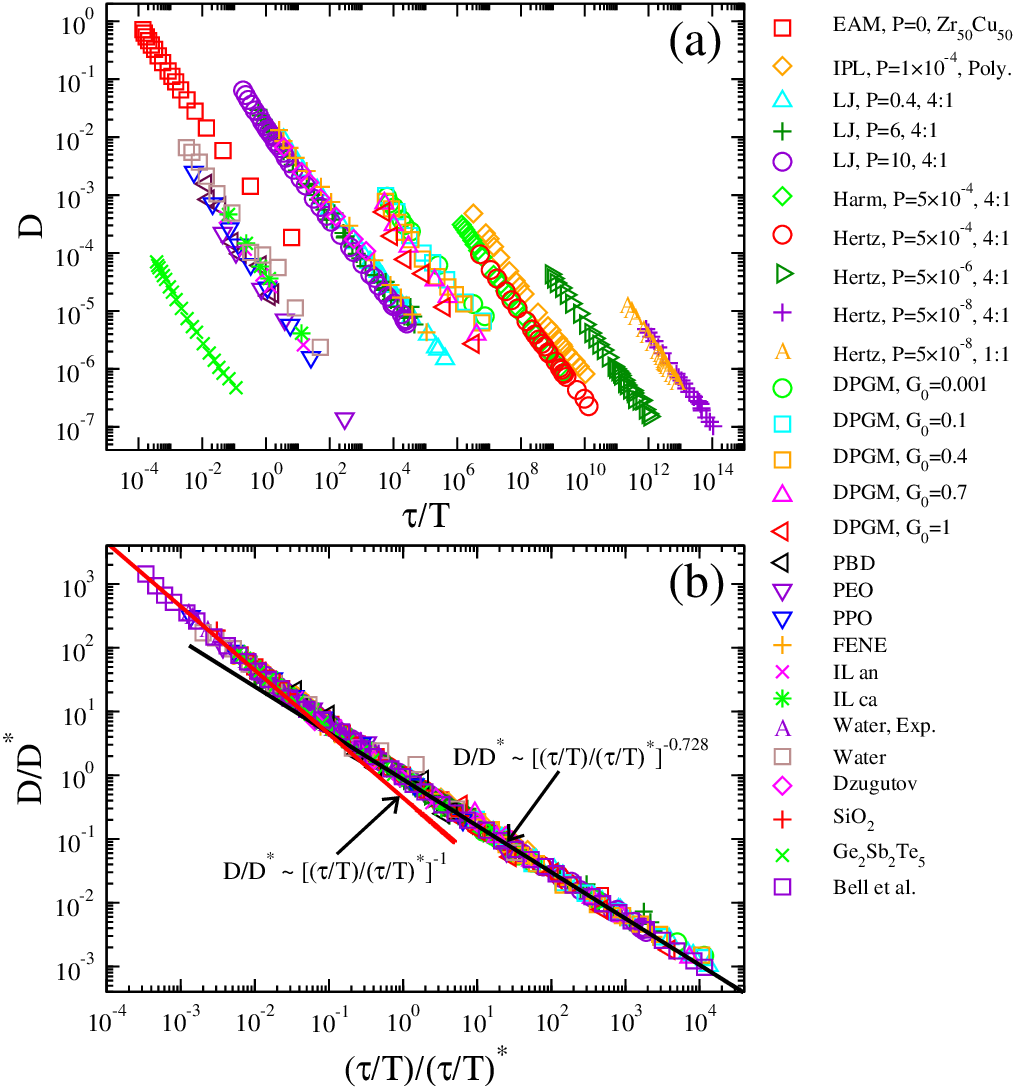}
\caption{\label{fig:SE} (a) Correlation between the self-diffusion coefficient $D$ and the ratio of structural relaxation time to temperature $\tau/T$. (b) Scaling collapse  between   $D$ and  $\tau/T$. The red line represents $D/D^*  \sim \left[(\tau/T)/(\tau/T)^*\right]^{-\xi}$ with $\xi=1$, whereas the black line corresponds to   $\xi=0.728$.  
In panel (b), all literature data have been rescaled by system-specific  constants so as to make them collapse  onto a single master curve. 
$(\tau/T)^*$ and $D^*$ for each system are listed in Table I in the Supplementary Information.
The literature data included are  supercooled water (Water)~\cite{Henritzi_2015_JCP}, polyethylene oxide ($\rm{PEO}$)~\cite{Henritzi_2015_JCP}, polypropylene oxide ($\rm {PPO}$)~\cite{Henritzi_2015_JCP}, polybutadiene ($\rm {PBD}$)~\cite{Henritzi_2015_JCP},  ionic liquids comprised of 1-n-butyl-3-methylimidazolium cations (IL ca) and hexafluorophosphate anions (IL an)~\cite{Henritzi_2015_JCP}, a polymer system featuring finite extensive nonlinear elastic  springs (FENE)~\cite{Starr_2013_JCP}, Dzugutov liquid (Dzugutov)~\cite{Gebremichael_2004_JCP},  a distinguishable-particle glassy model (DPGM)~\cite{Qin_2023_PRB} where the fragility (ranging from 26 to 343) is tuned by the parameter $\rm G_0$,  tetrahedral network ($\rm SiO_2$) \cite{Staley_2015_JCP}, and phase-change compound ($\text{Ge}_2\text{Sb}_2\text{Te}_5$)~\cite{Marcorini_2025_JCP}.   Additionally,  experimental data for supercooled bulk water (Water, Exp.)~\cite{Dehaoui_2015_PNAS} are included. 
We further extracted the data (violet squares) corresponding to the collapsed curve  for the dimensionless self-diffusion coefficient and the dimensionless viscosity, reported by Bell \textit{et al}.~\cite{Bell_2020_NC}.   All literature data, except where otherwise noted,  are  from simulation studies. Note that for certain extracted data,    only $D$ versus $\eta/T$ data are available in the original studies~\cite{Dehaoui_2015_PNAS,Staley_2015_JCP}. Hence, we incorporate these data under the assumption $ \tau \sim \eta$.
}
\end{figure}

\subsection*{Invariance of fractional SE relation exponent}

For supercooled liquids,  the SE relation holds at high temperatures above $T_{\rm{SEB}}$ but breaks down at low temperatures below $T_{\rm{SEB}}$.  
Note that   whether   $T_{\rm{SEB}}$  coincides with any established  characteristic temperature, such as the mode coupling temperature or  the onset temperature of slow dynamics~\cite{Berthier_2011_RMP},  remains a subject of debate~\cite{Berthier_2004_PRE, Flenner_2014_PRL,Sengupta_2013_JCP,Ediger_2000_AnnuRev,Staley_2015_JCP}.   The extent of the deviation from the SE relation has been numerically reported to be correlated with the fragility of supercooled liquids: more fragile supercooled liquids exhibit greater deviation~\cite{Adhikari_2021_JPCB,Ozawa_2016_JSM}.  This is consistent with several theoretical predictions~\cite{Pan_2005_PREjcp,Xia_2001_JPCBjcp}. However,  recent simulation studies suggest that this correlation is conditional~\cite{Qin_2023_PRB} or may be absent in certain systems~\cite{Bell_2020_NC}.
Moreover, whether there is a universal scaling between the self-diffusion coefficient and structural relaxation time remains uncertain, even in the simplest model supercooled liquids. 
For instance, Bell \textit{et al}.~\cite{Bell_2020_NC} identified a universal correlation between the self-diffusion coefficient and viscosity by employing an excess-entropy scaling framework in strongly correlating supercooled liquids.
However, the excess-entropy scaling approach used by Bell \textit{et al}. can not work  in weakly correlating systems~\cite{Pond_2011_SM,Krekelberg_2009_PRE,Gnan_2009_JCP}, which prevents the verification of a possible universal scaling in both strongly and weakly correlating supercooled liquids.

 Figure~\ref{fig:SE}(a) presents the dependence of $D$ on $\tau/T$ for all systems we studied, together with additional data compiled from the literature. Further details regarding the latter are provided in the caption of Fig.~\ref{fig:SE}. In particular, the literature data encompass supercooled $\rm SiO_2$, supercooled $\rm H_2O$, supercooled polymeric systems, and supercooled phase-change material. Note that our own data generated in this work comprise  strongly correlating liquids, such as those described by LJ and IPL interactions, as well as weakly correlating systems,  such as those governed by  Harm and Hertz potentials.
 It is evident that most curves in Fig.~\ref{fig:SE}(a) exhibit substantial differences. 
 
 By combining Eqs.~\ref{eq:D-alpha2max} and~\ref{eq:tauOverT-alpha2max}, one can obtain
\begin{equation}
\begin{split}
\label{eq:D-tauOverT}
D/D^* & = F^{-1}\{G\left[(\tau/T)/(\tau/T)^*\right]\} \\
&= H\left[(\tau/T)/(\tau/T)^*\right].
\end{split}
\end{equation}
This  indicates the existence of a universal scaling  between  the self-diffusion coefficient and structural relaxation time.    Such universal scaling is further verified  by the  collapse of  all our data  onto a  single master curve, as demonstrated in  Fig.~\ref{fig:SE}(b).  Notably,  the dynamical heterogeneity  acts as a critical bridge in the process of achieving such  scaling collapse.

Furthermore, we  rescale the literature data to ascertain whether they can also collapse onto the  master curve obtained using our own data in Fig.~\ref{fig:SE}(b). However, for the majority of $D$ versus $\tau/T$ data  available in the literature, corresponding values of $\alpha_{\rm 2,max}$ are unavailable. 
Accordingly, for each literature dataset, we simply rescale  $D$ by an appropriate constant $D^{*}$ and $\tau/T$ by  $(\tau/T)^{*}$  to maximize the collapsing degree of the corresponding data onto our master curve.  Notably, all  literature data extracted as well as our own  data generated can be superimposed onto the same master curve, as shown in Fig.~\ref{fig:SE}(b).  In particular,  the data corresponding to the collapsed curve for the dimensionless self-diffusion coefficient and dimensionless viscosity for strongly correlating systems, reported by Bell \textit{et al}.~\cite{Bell_2020_NC},  can likewise be mapped onto our master curve after rescaling by appropriate constants. However, further work is required to ascertain whether these scaling factors adopted  to collapse  literature data—particularly for complex supercooled liquids—can also be determined under the condition of constant dynamical heterogeneity. Should this not be the case, a subsequent challenge will be to elucidate  the physical meaning of these scaling factors.

In most numerical and experimental studies~\cite{Li_2019_PNAS, Ediger_2000_AnnuRev,Sengupta_2013_JCP,Adhikari_2021_JPCB,Debenedetti_2001_Nature,Staley_2015_JCP,Ediger_2000_AnnuRev,Glotzer_2000_JCP,Chen_2006_PNAS,Flenner_2013_JCP,Xu_2019_NP,Dehaoui_2015_PNAS,Marcorini_2025_JCP,Henritzi_2015_JCP,Swallen_2003_PRL,Becker_2006_PRL,Flenner_2014_PRL, Pan_2017_SR}, the fractional SE exponent  $\xi$    is determined by fitting the  $D$ versus $\tau/T$ (or its proxy) data at low temperatures near the glass transition. In this case, the choice of fitting range sometimes may significantly influence  the fitting outcome. We  also include  in   Fig.~\ref{fig:SE}(b)  the  literature data for several systems that  were reported to exhibit vastly different values of  $\xi$  in  corresponding original studies~\cite{Qin_2023_PRB,Henritzi_2015_JCP}. Notice that the variance of  $\xi$  therein was attributed to the change of fragility. Nevertheless, we observe that these data could collapse onto the same master curve after simply rescaling  $D$ and $\tau/T$ with appropriate constants, respectively. It is straightforward that rescaling  by constant factors  does not alter the value of  $\xi$ in any way.   Therefore, our analysis suggests that the fractional SE  exponent $\xi$  in all systems presented in   Fig.~\ref{fig:SE}(b) is approximately identical, and hence independent of  fragility  or other properties that may vary among these systems.  

Finally,  we will determine the functional form of $ H(x)$ in Eq.~\ref{eq:D-tauOverT}.  $H(x)$ can be described  piecewise by two power laws, i.e., $D/D^*  \sim \left[(\tau/T)/(\tau/T)^*\right]^{-\xi}$ with two different values of $\xi$.  Specifically, when $(\tau/T)/(\tau/T)^*$ is less than approximately 0.088, or $D/D^*$ exceeds approximately 5.048, $\xi=1$,  which  complies with the  SE relation. In contrast, when $(\tau/T)/(\tau/T)^*$ is greater than approximately 0.088, or $D/D^*$ is reduced below approximately 5.048, $\xi=0.728$,  thus indicating the emerge of the fractional SE relation.  The breakdown of the SE relation may start at different  temperatures across different supercooled liquids~\cite{Berthier_2004_PRE, Flenner_2014_PRL,Sengupta_2013_JCP,Ediger_2000_AnnuRev,Staley_2015_JCP}, yet according to our findings, it occurs at the  same value of $D/D^{*}$ or $(\tau/T)/(\tau/T)^*$, which varies with  the  fixed value of $\alpha_{\rm{2,max}}$ used to determine $D^{*}$ and $(\tau/T)^*$. 
Our results thus  suggest the invariance of the degree of  breakdown of the SE relation  across  various supercooled liquids studied. 
Nevertheless, it is important to emphasize that further investigations are required to evaluate the extent to which our findings can be generalized to a wider swath of supercooled liquids beyond the scope of the present work.

\section*{Discussion}

In summary, this study investigates the fractional SE relation from the   standpoint of dynamical heterogeneity. We demonstrate that imposing the condition of constant dynamical heterogeneity is essential for establishing a universal relationship between the self-diffusion coefficient and  structural relaxation time. Under this condition, system-specific characteristic scales associated with the self-diffusion coefficient and  structural relaxation time can be determined independently. The rescaled self-diffusion coefficient and the rescaled structural relaxation time, obtained using these characteristic scales, can each be represented as a univariate function of  dynamical heterogeneity. Consequently, a universal relationship between the self-diffusion coefficient and  structural relaxation time emerges via their mutual   dependence on dynamical heterogeneity.  

Bell \textit{et al}.~\cite{Bell_2020_NC}  derived a universal correlation between the self-diffusion coefficient and  viscosity by employing the excess-entropy scaling approach. However, the universality obtained from this approach is  conditional~\cite{Pond_2011_SM,Krekelberg_2009_PRE,Gnan_2009_JCP}. 
Although future work is required to  delineate  potential limitations of  the approach proposed in this work, we find the approach is demonstrably capable of capturing the behavior of specific classes of systems—such as those with  soft-core  potentials—that cannot be adequately described using the excess-entropy scaling approach~\cite{Pond_2011_SM}.

Moreover, we find there is a universal fractional SE relation with a constant fractional SE exponent among  all systems presented in this work. Depending on the specific classification method employed, our  systems examined span the spectrum from strong to fragile supercooled liquids~\cite{Angell_1991_JNCS,Angell_1995_Science}, or,  from strongly to weakly correlating supercooled liquids~\cite{Dyre_2014_JPCB}. Therefore, our finding challenges the dominant  view that the fractional SE exponent is correlated to fragility.   Note that while this work demonstrates the invariance of the fractional SE exponent across multiple distinct systems,  it does not elucidate the origin of the fractional SE relation.   

Finally, we seek to compare our  results with  experimental observations. In the following, we  concentrate on the case of supercooled water.
First, we notice that  two distinct fractional SE exponents were reported by Chen \textit{et al}.~\cite{Chen_2006_PNAS} in an experimental study of supercooled water confined in nanopores, which exhibits a fragile-to-strong transition~\cite{Geske_2016_AIP_slicaFTS,Ito_1999_Nature_waterFTS,Shi_2018_PANSfs}, i.e.,  a transition from fragile supercooled liquid  to strong one.  Specifically, an exponent of approximately 0.74 was reported in the higher-temperature regime where the system displays characteristics of a fragile liquid, whereas a smaller exponent of about 0.66 was obtained at lower temperatures where the system follows the behavior of a strong liquid. Although the exponent of 0.74 obtained in the fragile regime is close to our reported exponent of 0.728, the exponent of 0.66 in the strong regime exhibits a noticeable deviation from ours.  Accordingly, the findings reported in the present study cannot be directly generalized to describe the two fractional SE relations observed in the supercooled confined water by Chen \textit{et al}.~\cite{Chen_2006_PNAS}. 
An open question remains as to whether all supercooled liquids exhibiting a fragile-to-strong transition possess two distinct fractional SE exponents associated with the respective fragile and strong regimes~\cite{Geske_2016_AIP_slicaFTS,Chen_2006_PNAS}. If this is the case, a further issue of interest is whether these pairs of exponents are universal or system-dependent across different such supercooled liquids.  Moreover,
 one experimental study on supercooled bulk water by Dehaoui \textit{et al}.~\cite{Dehaoui_2015_PNAS} did not observe the existence of a fragile-to-strong transition, and we find that the fractional SE relation reported in their study could be well fitted by the master curve presented  in Fig.~\ref{fig:SE}(b).

\section*{Methods}

The interaction potentials adopted in this work include the soft-core potential~\cite{Wang_2012_SM}, the Lennard-Jones potential~\cite{Kob_1994_PRL},  the inverse-power-law  potential~\cite{Ninarello_2017_PRX}, and the realistic embedded atom method potential~\cite{Cheng_2009_EAM}.

The soft-core potential is given by
\[
U(r_{ij})=
\begin{cases}
\displaystyle \frac{\varepsilon_{ij} \big(1-r_{ij}/\sigma_{ij}\big)^\alpha}{\alpha}, & r_{ij}\le \sigma_{ij} \\[6pt]
0, & r_{ij}> \sigma_{ij}
\end{cases}
\]
where $r_{ij}$ denotes the distance between particle $i$ and particle $j$.  The harmonic potential (Harm) corresponds to $\alpha=2$, whereas the Hertzian potential (Hertz) corresponds to $\alpha=2.5$. The studied systems  consist of two types of particles, denoted as type A and type B, with identical mass. We consider two distinct particle-number ratios between component A and component B, namely, \(N_\mathrm{A}:N_\mathrm{B}=4:1\) and \(N_\mathrm{A}:N_\mathrm{B}=1:1\). In the case of the \(N_\mathrm{A}:N_\mathrm{B}=4:1\) mixture, we set  $\epsilon_\mathrm{AB}=1.5\epsilon_\mathrm{AA}$, $\epsilon_\mathrm{BB}=0.5\epsilon_\mathrm{AA}$, $\sigma_\mathrm{AB}=0.8\sigma_\mathrm{AA}$, and $\sigma_\mathrm{BB}=0.88\sigma_\mathrm{AA}$.
For   systems with $N_\mathrm{A}:N_\mathrm{B}=1:1$,  we set $\epsilon_\mathrm{AB}=\epsilon_\mathrm{BB}=\epsilon_\mathrm{AA}$, $\sigma_\mathrm{AB}=\frac{6}{7}\sigma_\mathrm{AA}$, and $\sigma_\mathrm{BB}=\frac{5}{7}\sigma_\mathrm{AA}$.

The Lennard-Jones (LJ) potential reads
\[
U(r_{ij})=
\begin{cases}
   f(r_{ij}) - f(r_{ij}^{c}) - (r_{ij} - r_{ij}^{c}) f'(r_{ij}^{c}), & r_{ij}\le r_{ij}^c \\[6pt]
0, & r_{ij}> r_{ij}^c
\end{cases}
\]
where $f(r_{ij})=4\varepsilon_{ij}\left[\left(\dfrac{\sigma_{ij}}{r_{ij}}\right)^{12}-\left(\dfrac{\sigma_{ij}}{r_{ij}}\right)^6\right]$,  and the potential cutoff distance $r_{ij}^c = 2.5\sigma_{ij}$.  We study binary LJ mixtures with $N_\mathrm{A}:N_\mathrm{B}=4:1$.  For all  LJ systems studied~\cite{Kob_1994_PRL}, we set $\epsilon_\mathrm{AB}=1.5\epsilon_\mathrm{AA}$, $\epsilon_\mathrm{BB}=0.5\epsilon_\mathrm{AA}$, $\sigma_\mathrm{AB}=0.8\sigma_\mathrm{AA}$, and $\sigma_\mathrm{BB}=0.88\sigma_\mathrm{AA}$.  All particles have the same mass.

The inverse-power-law (IPL) potential takes the form
\[
U(r_{ij})=
\begin{cases}
\left( \frac{\sigma_{ij}}{r_{ij}} \right)^{12} +  c_{0} + c_{2} \left( \frac{r_{ij}}{\sigma_{ij}} \right)^{2} + c_{4} \left( \frac{r_{ij}}{\sigma_{ij}} \right)^{4}, & r_{ij}\le r_{ij}^c \\[6pt]
0, & r_{ij}> r_{ij}^c
\end{cases}
\]
where we  set  $r_{ij}^c = 1.25\sigma_{ij}$, and the constants \(c_0\), \(c_2\), and \(c_4\) are chosen such that  \(U(r_{ij})\) and its first and second derivatives are continuous at  \(r_{ij}^c\). We study  polydisperse  systems characterized by a particle diameter distribution given by \(P(\sigma) \sim \frac{1}{\sigma^{3}}\), where the diameter   \(\sigma\)  varies continuously within the interval \(0.73 \leq \sigma \leq 1.62\).  We set the cross-diameter according to the non-additive rule,  $\sigma_{ij} = \frac{\sigma_i + \sigma_j}{2} \left( 1 - 0.2 \lvert \sigma_i - \sigma_j \rvert \right)$.   All particles have the same mass.

Additionally, we  study model  $\mathrm{Zr}_{50}\mathrm{Cu}_{50}$ metallic supercooled liquids, where the interatomic interactions  are described by the realistic embedded atom method (EAM) potential~\cite{Cheng_2009_EAM}. 

We perform  molecular dynamics  simulations, using LAMMPS~\cite{lammps},   in three-dimensional systems under the \textit{NPT} ensemble. Periodic boundary conditions are applied in all  directions throughout the simulations. 
The integration time step adopted in  simulations ranges from 0.001 to 0.01, and is adjusted based on the  temperature and pressure. Simulation data are collected  after each system reaches full equilibration, with the equilibration duration set to no less than ten times the temperature- and pressure-dependent structural relaxation time $\tau$. The duration of the production simulation run is also set to no less than ten times $\tau$.  
Although the majority of our results presented  are obtained from systems composed of  $N = 1000$ particles (with the exception of $N = 8000$  particles for $\mathrm{Zr}_{50}\mathrm{Cu}_{50}$ systems), we have  partially verified that our conclusions are not  affected by finite-size effects.

We calculate the self-intermediate scattering function~\cite{Kob_1994_PRL} defined as 
\[
F_s(k,t) = \frac{1}{N} \left\langle \sum_{j=1}^{N} \exp\left[ i\vec{k}\cdot\big(\vec{r}_j(t)-\vec{r}_j(0)\big) \right] \right\rangle,
\]
where $\vec{r}_j(t)$ denotes the position of particle $j$ at time $t$, and the magnitude $|\vec{k}|$ is chosen approximately at the first peak of the static structure factor~\cite{Allen_1987}. The notation $\langle \cdot \rangle$ represents the time average. The structural relaxation time $\tau$ is determined such that  $F(k,\tau)=1/\mathrm{e}$.

We  calculate the mean squared displacement (MSD)
\[
\Delta^2(t) = \left\langle \frac{1}{N}\sum_{i=1}^{N} \big|\vec{r}_i(t) - \vec{r}_i(0)\big|^2 \right\rangle.
\]
The self-diffusion coefficient $D$ is determined by fitting the long-time MSD to the Einstein relation, 
\[
D = \lim_{t\to\infty} \frac{\Delta^2(t)}{6t}.
\]

The quantities  commonly used to characterize dynamic heterogeneity include the non-Gaussian parameter and the four-point dynamical susceptibility~\cite{Glotzer_2000_JCP,Kim_2013_JCP,Kob_1997_PRL_A2,Henritzi_2015_JCP,Starr_2013_JCP,Adhikari_2021_JPCB,Sengupta_2013_JCP}. 

The non-Gaussian parameter $\alpha_2(t)$ is defined as
\[
\alpha_2(t) = \frac{3\left\langle \big|\vec{\Delta r}(t)\big|^4 \right\rangle}{5\left\langle \big|\vec{\Delta r}(t)\big|^2 \right\rangle^2} - 1,
\]
where $\vec{\Delta r}(t) = \vec{r}(t) - \vec{r}(0)$ represents the particle displacement  within time $t$.
In supercooled liquids, $\alpha_2(t)$ typically exhibits a non-monotonic dependence on time $t$ at each temperature. We adopt the maximum value of $\alpha_2(t)$, denoted as $\alpha_{2,\mathrm{max}}$, to quantify the strength of dynamical heterogeneity.  

The four-point dynamical susceptibility $\chi_4(t)$ measures the variation of an overlap function $Q(a,t)$, and takes the form
\[
\chi_4(a,t) = \frac{1}{N} \left[ \left\langle Q(a,t)^2 \right\rangle - \left\langle Q(a,t) \right\rangle^2 \right],
\]
where  
\[
Q(a,t) = \sum_{i=1}^{N} W_a\left( \big|\vec{r}_i(t) - \vec{r}_i(0)\big| \right)
\]
 quantifies the similarity between two configurations separated by a time interval $t$.  $W_a(x) = 1$ if $x \leq a$, otherwise, $W_a(x) = 0$.   Here, we set $a=0.3$.  Similar to the non-monotonic time dependence observed for $\alpha_2(t)$  in supercooled liquids, $\chi_4(t)$  exhibits as well a non-monotonic dependence on time. We adopt the peak value of $\chi_4(t)$, denoted as $\chi_{4,\text{max}}$, to quantify the magnitude of dynamical heterogeneity.  The results obtained from the analysis of $\chi_{4,\text{max}}$ are provided in the Supplementary Information.

\section*{ Acknowledgments}

\noindent We thank one anonymous referee of our previous work for drawing our attention to  Ref.~\cite{Bell_2020_NC}.    
 We acknowledge  the support from   National Natural Science Foundation of China (Grant Nos. 12522503 and 12374202) and Anhui Projects (Grant Nos. 2022AH020009, S020218016 and Z010118169).  We  also acknowledge Hefei Advanced Computing Center, Beijing Super Cloud Computing Center, and the High-Performance Computing Platform of Anhui University for providing computing resources.

\end{document}